\documentclass{article}
\RequirePackage[a4paper]{geometry}
\usepackage{mathptmx,graphicx}
\usepackage{siunitx}
\usepackage{amsmath}
\usepackage{amssymb}
\usepackage{amsthm}
\usepackage{abstract}
\usepackage{subfig}
\usepackage{subcaption}

\usepackage[
backend=bibtex,
style=nature,
]{biblatex}
\begin{document}

\begin{center}
{\Large\bfseries Nonlinear dynamics of femtosecond laser interaction with the central nervous system in zebrafish \par}
\vspace{3ex}
 {\bfseries Soyeon Jun$^{1,2,3}$, Andreas Herbst$^{1,2}$, Kilian Scheffter$^{1,2}$, Nora John$^{1,4,5}$, Julia Kolb$^{1,4,5}$, Daniel Wehner$^{1,4,*}$, Hanieh Fattahi$^{1,2,*}$\par}
    {\footnotesize\itshape
    1. Max Planck Institute for the Science of Light, 91058 Erlangen, Germany\\
    2. Friedrich-Alexander University Erlangen-Nürnberg, 91058 Erlangen, Germany\\
    3. Friedrich-Alexander-Universität Erlangen-Nürnberg (FAU), Erlangen Graduate School in Advanced Optical Technologies (SAOT), 91052 Erlangen, Germany \\
    4. Max-Planck-Zentrum für Physik und Medizin, 91058 Erlangen, Germany\\
    5. Department of Biology, Animal Physiology, Friedrich-Alexander-University Erlangen-Nürnberg, 91058 Erlangen, Germany\\
    * Correspondence: hanieh.fattahi@mpl.mpg.de, daniel.wehner@mpl.mpg.de
    \par}
     \vspace{3ex}
\end{center}

\begin{abstract}
Understanding the photodamage mechanism underlying the highly nonlinear dynamic of femtosecond laser pulses at the second transparent window of tissue is crucial for label-free microscopy. Here, we report the identification of two cavitation regimes from 1030\,nm pulses when interacting with the central nervous system in zebrafish. We show that at low repetition rates, the damage is confined due to plasma-based ablation and sudden local temperature rise. At high repetition rates, the damage becomes collateral due to plasma-mediated photochemistry. Furthermore, we investigate the role of fluorescence labels with linear and nonlinear absorption pathways in optical breakdown. To verify our findings, we examined cell death and cellular responses to tissue damage, including the recruitment of fibroblasts and immune cells post-irradiation. These findings contribute to advancing the emerging nonlinear optical microscopy techniques and provide a strategy for inducing precise, and localized injuries using near-infrared femtosecond laser pulses.
\end{abstract}

\section*{Introduction}

Light microscopy has revolutionized our ability to observe and understand biological processes. However, does the use of intense light irreversibly perturb cellular processes? What is the most accurate definition of biological damage caused by our observation? Does the instantaneous death of irradiated cells, as noted by \cite{talone2021phototoxicity}, constitute damage? Or should any reversible cellular impairment in a living organism already be classified as damage \cite{zhang2022phototoxic}? What if the illumination leads to cell apoptosis by slowing down vital metabolic processes by degrading intracellular organic compounds \cite{wagner2010light}? 

Life on Earth depends on light and is adapted to the solar flux of less than \SI{1.4}{kW/m^2}. It has been argued that this value should be used as a safe irradiance reference when observing dynamic biological processes \cite{stelzer2015light}. However, the sun is a continuous source of incoherent radiation with a non-uniform multi-octave spectral coverage. While the organisms are adopted to harvest the energy of frequencies with the higher spectral density of the sun for their development through photosynthesis, they stayed ignorant of other lower spectral density components present in the solar spectrum \cite{zhen2022photosynthesis}. Furthermore, the photodamage effect of continuous light on biological specimens differs from that of pulsed lasers at similar photon flux \cite{konig1999pulse,KOESTER19992226, konig2000laser}. Therefore, it is crucial to carefully consider these factors when light microscopy is used for imaging and in medical applications. In the past few decades, two significant types of microscopy have gained widespread usage: fluorescence microscopy, which utilizes the emission of linearly or non-linearly excited fluorophores to capture microscopic images of the sample \cite{denk1990two,kerr2008imaging, horton2013vivo,ANDRESEN200954, Velasco:21, CALOVI2021105420}; and label-free microscopy, which relies on the emission of the molecular composition of the sample caused by an induced nonlinear polarization at high laser peak intensities \cite{helmchen2005deep,weigelin2016third, squier1998third, witte2011label, debarre2014mitigating, ji2013rapid, muller19983d, barad1997nonlinear,doi101021jp035693v, barad1997nonlinear}. Despite the distinct imaging methodologies utilized in these two categories, they share a common mechanism of photodamage.

\begin{figure}[h!]
            \centering
            \includegraphics[width=1\textwidth]{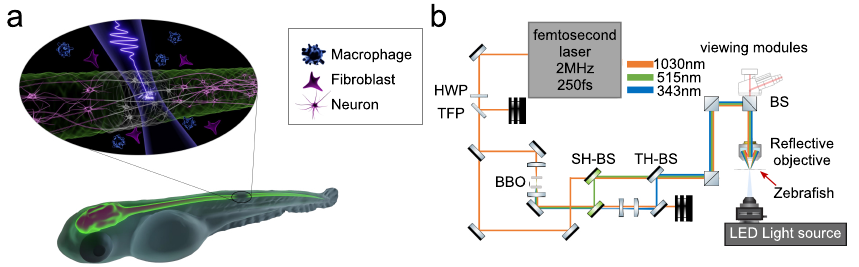}
            \caption{\textbf{Scheme of the experimental procedure.} a) Zebrafish larvae were used as a model organism to investigate the \textit{in vivo} interaction of femtosecond pulses with tissues. Precisely targeted femtosecond laser pulses were focused on the central nervous system (CNS) under various irradiation settings. The cellular dynamics resulting from this pulse-tissue interaction were meticulously observed over time, focusing on neural cells and gauging the responsiveness of macrophages and fibroblasts. b) Illumination setup for studying the interaction of femtosecond pulses at various wavelengths with the zebrafish CNS. HWP: Half Wave Plate; TFP: Thin Film Polarizer; SH-BS: Second Harmonic Beam Splitter; TH-BS: Third Harmonic Beam Splitter; BBO: Beta Barium Borate crystal; BS: Beam Splitter.}
            \label{fig1}
\end{figure}
When a pulsed laser interacts with tissue, the molecules in the sample get excited to a higher energy state through single-photon or multiphoton absorption, depending on the illumination peak intensity and optical frequency. The molecules can then return to the ground state either non-radiatively, leading to thermal damage, or via de-excitation processes. De-excitation involves dissociation or changes in the redox state of the absorbing molecule, energy transfer to a dioxygen molecule, or formation of reactive oxygen species (ROS), resulting in localized chemical perturbations and cell death \cite{tosheva2020between,tirlapur2001femtosecond, khan2015molecular, dillenburg2014laser}. At extremely high peak intensities, ionization of molecules and formation of a low-density plasma takes place, leading to photochemical damage or ablation and production of supersonic shock waves, which likewise lead to tissue damage \cite{olivie2008wavelength, sarpe2006plasma,astafiev2023femtosecond, vogel2002femtosecond}. In both of these categories of microscopy, the interaction between the laser pulses and tissue involves absorbers, predominantly consisting of water molecules 
\cite{linz2016wavelength} or other tissue components depending on the excitation wavelength \cite{hemmer2016exploiting, xu2021nir, hong2017near, smith2009second, weissleder2001clearer} in label-free microscopy, or fluorescence labels in fluorescence microscopy \cite{laissue2017assessing, waldchen2015light}. While ROS generation significantly impacts cell damage in fluorescence microscopy, its contribution in label-free microscopy is negligible. Therefore, there has been significant attention on the use of different label-free microscopy techniques in the recent years \cite{olivier2010cell, ferrer2023label, sun2004higher, oron2004depth, olivier2010cell, weigelin2016third, olivier2010cell, andresen2009infrared}. Moreover, due to its large penetration depth and image contrast, label-free microscopy has become a crucial tool for deep-tissue and long-range imaging. The complementary information provided by both categories of microscopy has led to the development of more advanced techniques, such as multimodal nonlinear microscopy \cite{Wu:22, you2018intravital, clark2022pulse, matsui2022label}.

Over the last decades, a significant body of research has been conducted on the effects of short laser pulses on thin \textit{in vitro} and \textit{ex vivo} samples in nonlinear fluorescence microscopy \cite{chen2002wavelength, MAGIDSON2013545, Fu:06, debarre2014mitigating, talone2021phototoxicity, Maioli:20}. Despite these numerous investigations, a comprehensive study on how laser irradiation affects cell viability in a long-range imaging setting \textit{in vivo}, specifically for label-free and multimodal nonlinear microscopy, has not been fully explored. This study leverages the vertebrate species zebrafish (\textit{Danio rerio}) to delve into the mechanisms of photodamage in tissue at a cellular level triggered by femtosecond excitation pulses. Our objectives revolve around elucidating several key questions: How do the dynamics of different photodamage mechanisms unfold across a spectrum of femtosecond pulses, particularly at near-infrared (NIR)? What are the dynamics behind the photodamage at high peak intensities at NIR, and does the repetition rate of the laser pulses influence the driving mechanism? Is it feasible to reach plasma formation without inducing gradual damage to the sample? What is the role of thermal effects in photodamage at high peak intensities? To what extent does the ROS generation due to fluorescence proteins cause acceleration in photodamage? How does the damage threshold scale relative to the numerical aperture of focusing optics?

\section*{Results}
\begin{figure}
            \centering
            \includegraphics[width=1\textwidth]{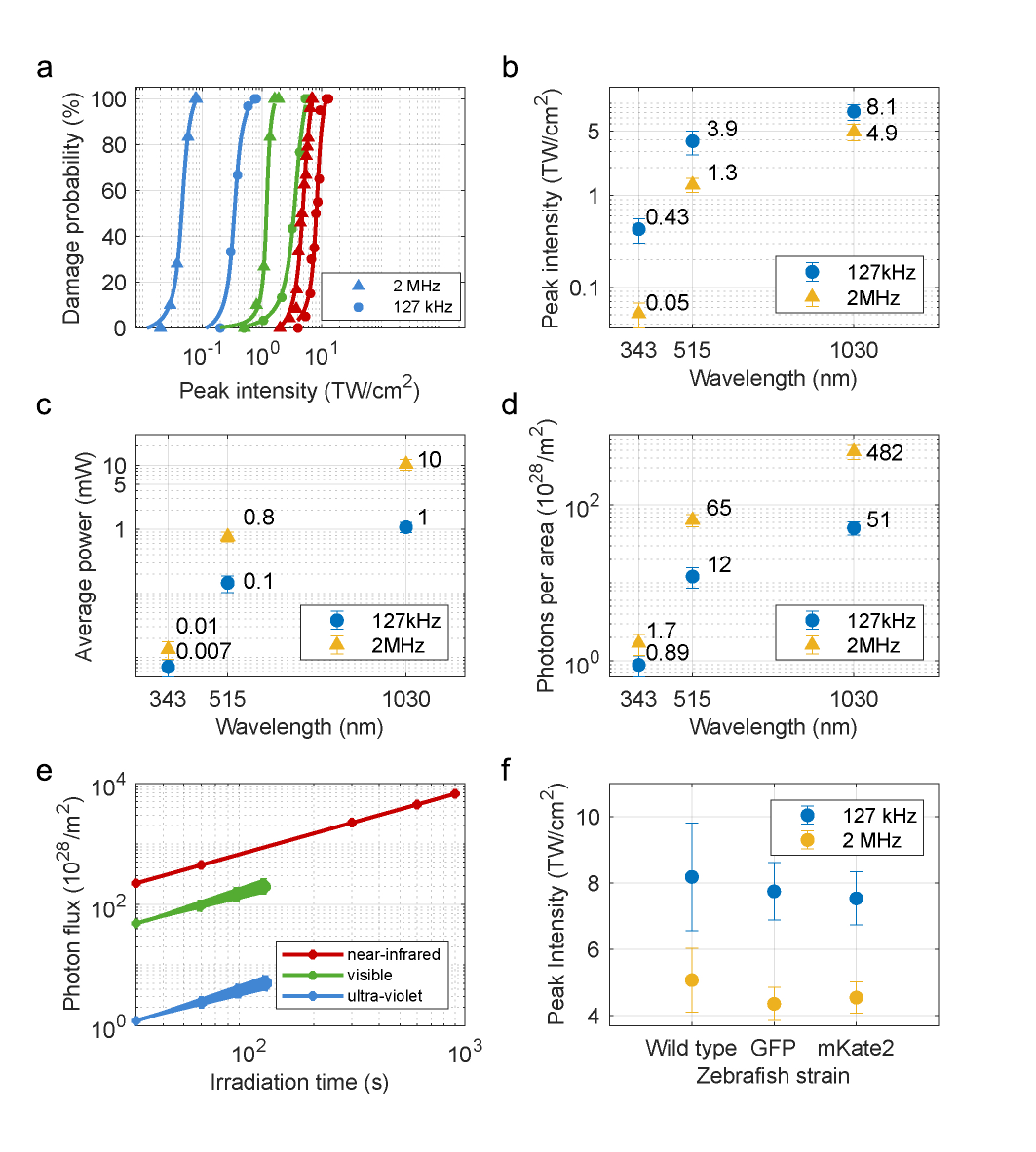}
            \caption{\textbf{Femtosecond laser interaction with the central nervous system of zebrafish.} a) The damage probability of the zebrafish spinal cord versus peak intensity of 250\,fs pulses at various irradiation wavelengths and laser repetition rates. Irradiation at 343\,nm, 515\,nm, and 1030\,nm are represented by blue, green, and red curves, respectively. The circle indicates irradiation at 127\,kHz repetition rates, while the triangle shows irradiation at 2\,MHz repetition rates. b) The average values of the peak intensity threshold at various wavelengths and laser repetition rates.  c) The average power damage threshold at various wavelengths and laser repetition rates. d) The comparison of the average photons per area threshold for different parameters. Dwell time was fixed to 30\,s. e) The damage of the spinal cord was observed when the samples were irradiated at 343\,nm and 515\,nm below the peak intensity damage threshold and at longer dwell times of 30\,s and 60\,s respectively. For below-threshold irradiation at 1030\,nm, no damage was observed by increasing the irradiation beyond 900\,s. The thickness of the curves for 343\,nm, and 515\,nm illustrates the severity of the damage qualitatively. f) The average values of the peak intensity threshold at 1030\,nm and various laser repetition rates for wild-type (non-labeled) zebrafish and zebrafish in which neurons were labeled by either mKate2 (assessed in \textit{elavl3}:rasmKate2 transgenic animals) or GFP (assessed in \textit{elavl3}:GFP-F transgenic animals). b-d,f) The error bars show standard deviation.}
            \label{fig2}
\end{figure}

Evaluating \textit{in vivo} phototoxicity in tissues is considerably more intricate than in cultured cells, where probes for ROS production or observation of cell death through carbonization are common. In this study, we used zebrafish larvae at three days post-fertilization (dpf) to investigate interactions between femtosecond pulses and tissues in a living organism. We employed two complementary criteria to assess photodamage at the level of the spinal cord and surrounding trunk tissue. Short-term damage was evaluated by monitoring loss of tissue integrity at the whole tissue level and by assessing specific cell types (neurons, glial cells, keratinocytes), which were monitored through fluorescence labels in specific transgenic zebrafish lines. Long-term damage was assessed by observing axon regeneration over a period of two days, histological stains to detect cell death, and injury-associated recruitment of reactive fibroblasts and immune cells (macrophages) within 24 hours post-irradiation (hpi) \cite{tsata2021switch,tsarouchasdynamic}.

Figure \ref{fig1} a and b present a visual summary of the experiments conducted to investigate the damage mechanism upon the interaction of femtosecond laser pulses with the central nervous system (CNS) of a zebrafish larva. To generate the required femtosecond laser pulses at various wavelengths for this study, the output of a ytterbium amplifier delivering 250\,fs pulses at 1030\,nm was up-converted to 515\,nm and 343\,nm through second-harmonic and third-harmonic generation, respectively. A Pockels cell was utilized for pulse picking and reducing the laser pulse train's repetition rate from 2\,MHz to 127\,kHz, enabling the differentiation of thermal damage contributions across various irradiation regimes. The generated femtosecond laser pulses were focused on the spinal cord using a reflective microscope objective. An imaging system was developed to monitor and adjust the focus position across different sections of the zebrafish. To monitor tissue damage following the different irradiation paradigms, we employed different transgenic fluorescence reporter zebrafish lines to examine the photodamage of the skin and spinal cord.

Two measurements were conducted to differentiate photodamage from linear processes, such as one-photon absorption and thermal damage (which should scale linearly with dwell time and photon flux), from nonlinear processes triggered by laser peak intensity. In the first set of measurements, samples were exposed to varying peak intensities while maintaining fixed dwell times. In the second series, laser pulses with varying dwell times and tissue at constant pulse energy were examined. Here, the peak intensity threshold was kept below the average damage intensity threshold measured in the previous experiment. Additionally, these measurements were carried out at two distinct repetition rates to discern the underlying photodamage processes further. 

In the first series of measurements, a total number of 180 zebrafish larvae were individually irradiated for 30\,s at various peak intensities, repetition rates, and irradiation wavelengths. Glial cells and neurons of the spinal cord were labeled with the fluorescence proteins GFP and mKate2, respectively (assessed in \textit{her4.3}:GFP-F; \textit{elavl3}:rasmKate2 double transgenic animals). Figure \ref{fig2} a displays the spinal cord damage probability versus peak intensity, while Figure \ref{fig2} b presents the average damage intensity threshold for different parameters. Our data indicate the nonlinear scaling of the damage threshold versus irradiation wavelength. Furthermore, irradiation at a lower repetition rate across all wavelengths leads to a higher damage threshold than irradiation at higher repetition rates within the same dwell time. It has been observed that the damage behavior for the 1030\,nm irradiation has a different qualitative behavior compared to 515\,nm and 343\,nm. The lesion's dimension expands as the peak intensity increases for the two shorter irradiation wavelengths. However, damage occurs abruptly for the 1030 nm wavelength, immediately generating visible mechanical shockwaves. Figure \ref{fig2} c shows the corresponding average power threshold for damage at various wavelengths and repetition rates. Figure \ref{fig2} d summarizes the photons per area related to two irradiation regimes (see Supplementary Figure 1 for more information). 

The dwell time in the second series of measurements increased while the peak intensity was below the damage threshold. Our study on photon flux revealed that, for irradiation at 343\,nm and 515\,nm, the lesion expands with increased dwell time, showing gradual damage. However, the response is entirely different for 1030\,nm irradiation. No damage was observed by increasing the dwell time at a peak intensity below the damage threshold, as judged by loss of fluorescence signal or compromised tissue integrity. Figure \ref{fig2} e shows photon flux versus dwell time for three different wavelengths. At 343\,nm, photo damage is visible following 30\,s of irradiation, signifying that linear mechanisms predominantly drive tissue damage at this wavelength due to absorption by DNA \cite{mohanty2002comet, he2009mechanism, rapp2013soft} (see Supplementary Figure 2). In contrast, at 515\,nm irradiation wavelength, the damage was observed at one order of magnitude higher photon flux threshold, hinting at the involvement of nonlinear processes. At 1030\,nm, it is possible to extend the irradiation duration to several minutes. 

Further measurements were performed on animals subjected to irradiation at 1030\,nm as they showed distinct behaviors compared to irradiation at 343\,nm and 515\,nm. The mKate2 fluorophore has a high absorption cross-section for two-photon absorption at 1030\,nm, while the nonlinear absorption cross-section of the GFP fluorophore at 1030\,nm is weaker \cite{lambert2019fpbase, drobizhev2011two}. Therefore, we used transgenic lines in which all neurons were either labeled with GFP (assessed in \textit{elavl3}:GFP-F transgenic zebrafish) or mKate2 (assessed in \textit{elavl3}:rasmKate2 transgenic zebrafish) to isolate the additional role of the fluorescent labels in photodamage. Figure \ref{fig2} f compares the averaged peak intensity threshold for the two scenarios in addition to the wild-type fish. The damage threshold of samples labeled by GFP is higher than mKate2 labeled samples for irradiation at 127\,kHz, indicating the role of the two-photon absorption pathway and ROS generation of mKate2. However, at 2\ MHz repetition rates, the difference between the damage threshold at both labels becomes marginal, indicating that at these regimes, the role of the plasma-mediated chemical effects and nonlinear photochemistry becomes prominent (see Supplementary Figure 1).

\begin{figure}[h!]
    \centering
    \includegraphics[width=1\textwidth]{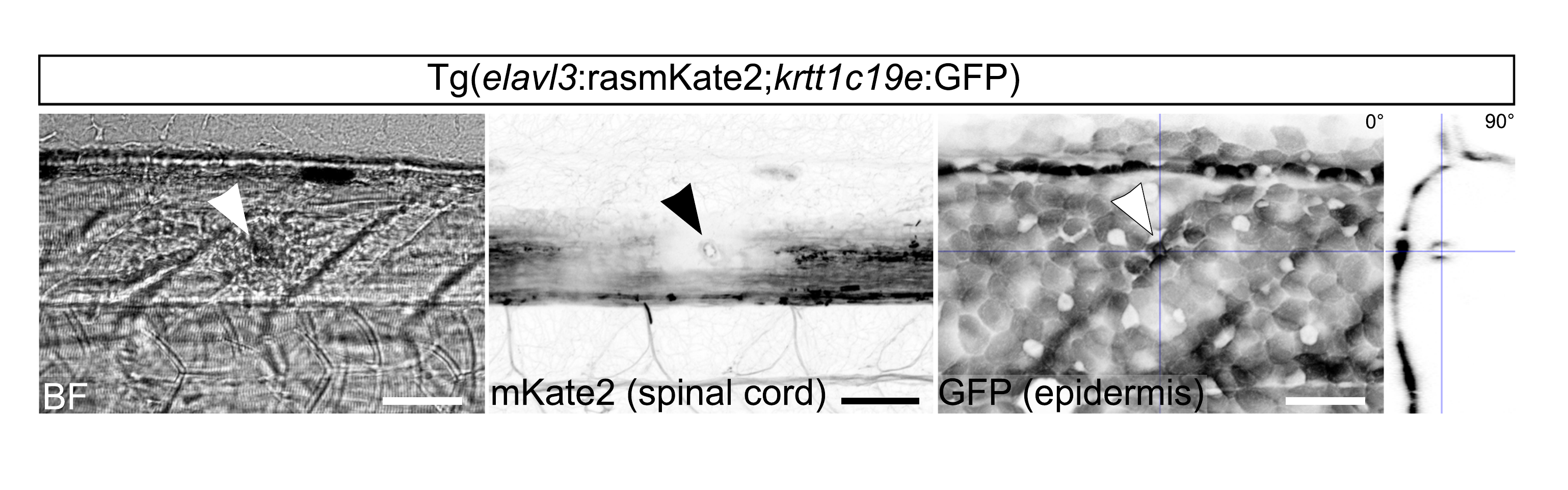}
    \caption{\textbf{Tissue damage is confined to the focal area.} Trunk of an \textit{elavl3}:rasmKate2;\textit{krtt1c19e}:EGFP double transgenic zebrafish larva with the fluorescently labeled spinal cord (neurons; mKate2) and epidermis (basal keratinocytes; GFP) after 30\,s irradiation with 1030 nm pulses above the damage threshold. Arrowheads indicate the focus of irradiation. While damage is visible at the level of the spinal cord, the superficially located epidermis remains intact. Images shown are bright-field (BF) recordings and orthogonal projections of the confocal image (xy or yz view). Scale bars: \SI{50}{\micro\metre}.}
    \label{fig2_a}
\end{figure}

For the samples deemed healthy upon irradiation of intense femtosecond pulses at 1030\,nm, we examined potential damage to the skin using (\textit{krtt1c19e}:GFP) transgenic zebrafish with fluorescently labeled basal keratinocytes of the bilayered epidermis. Figure \ref{fig2_a} depicts the bright-field, GFP-labeled basal keratinocytes and mKate2-labeled neurons (visualized in \textit{elavl3}:rasmKate2 transgenic animals) following irradiation above the damage intensity threshold at 1030\,nm. After irradiation, we observed a sharply declined fluorescence of the neuronal marker but not the epidermal marker, indicating that tissue damage was largely limited to the focal area. Similar results are observed for irradiation at the other two wavelengths, where no apparent damage was detected on the epidermis (see Supplementary Figure 3 for more information).

\begin{figure}[h!]
    \centering
    \includegraphics[width=1\textwidth]{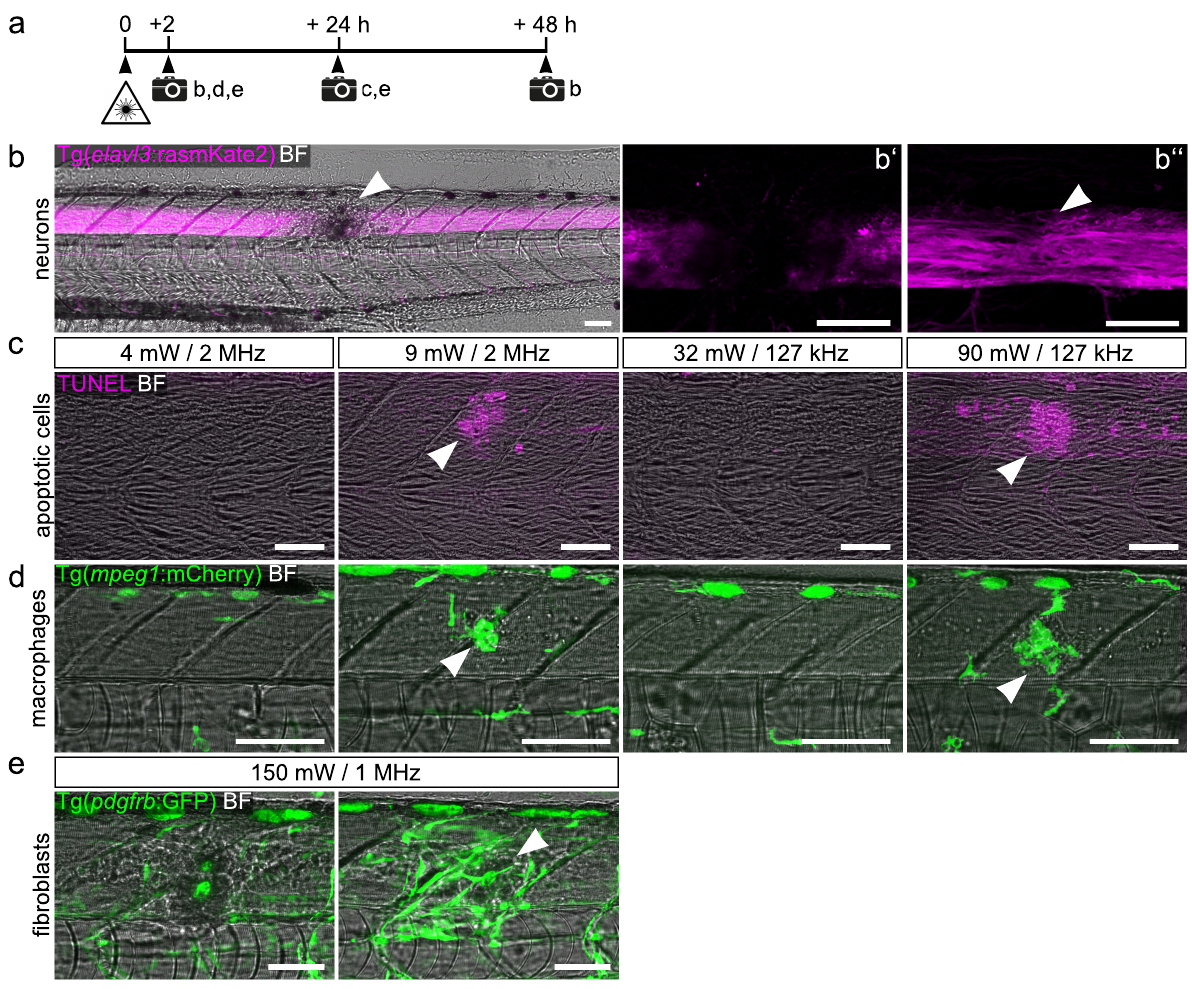}
    \caption{\textbf{Cellular responses to different irradiation paradigms below and above the damage threshold at 1030\,nm.} a) The timeline of experiments. Timepoints of irradiation and analysis are indicated. b) Spinal cord lesion induced by 30\,mW pulses at 2\,MHz (arrowhead in overview). The same \textit{elavl3}:rasmKate2 transgenic animal is shown at 2\,hpi ($b^{'}$) and 48\,hpi ($b^{''}$). Note regrowth of white matter tracts at 48\,hpi (arrowhead in b). The neurons of the zebrafish spinal cord are genetically labeled with mKate2 fluorescent protein. c) Cell death (arrowheads), as detected by the TUNEL assay, is observed in the irradiation site at above-damage threshold peak intensities but not below the damage threshold intensities. d) Recruitment of macrophages (arrowheads) to the irradiation site is observed at 16\,mW at 2\,MHz and 1.6\,mW at 127\,kHz; however, not at below damage threshold irradiations at the average powers of 5.8\,mW at 2\,MHz and 0.7\,mW at 127\,kHz (visualized with \textit{mpeg1}:Cherry transgenic animals). e) Irradiation at above damage threshold intensity leads to the recruitment of reactive fibroblasts in the irradiation site (arrowhead; visualized with \textit{pdgfrb}:GFP transgenic animals). b-e) Scale bars: \SI{50}{\micro\metre}.}
    \label{fig3}
\end{figure}

To exclude the possibility that the reduction of fluorescence signal in the spinal cord is a consequence of bleaching of the fluorophore and not due to tissue damage, we reassessed the irradiated animals at 24\,hpi and 48\,hpi. This showed an imperfect fluorescence pattern of the white matter tracts, supporting the regrowth of irradiation-damaged axonal fibers rather than recovery of fluorescence, which occurs in larval zebrafish within 48 hours after mechanical transection as previously reported \cite{wehner2017wnt} (Figure \ref{fig3} a,b). To provide further evidence for tissue damage or its absence in the respective conditions, we assessed cellular apoptosis by TUNEL assay and monitored the recruitment of immune cells and fibroblasts (Figure \ref{fig3} c-e) to the irradiated site \cite{tsata2021switch,tsarouchasdynamic}. Consistent with tissue damage, we observed the emergence of TUNEL\textsuperscript{+} cells as well as the recruitment of reactive fibroblasts (visualized with \textit{pdgfrb}:GFP transgenic animals) and macrophages (visualized with \textit{mpeg1}:mCherry transgenic animals) in samples irradiated at 1030\,nm for 30\,s above the damage threshold. In contrast, for samples irradiated just below the damage threshold, we did not detect TUNEL signal, fibroblasts, or macrophages accumulating at the irradiation site. This supports our observation of the highly nonlinear scaling of the damage at NIR and the absence of any gradual damage at this wavelength.

\section*{Discussion}

The photodamage peak intensity threshold for CNS tissue of zebrafish when irradiated by femtosecond pulses at 1030\,nm is higher compared to irradiation at its harmonics. It was observed that when samples are subjected to femtosecond pulses at 343\,nm and 515\,nm, the extent of the photodamage is gradual and scales proportional to irradiation photon flux. A lower ablation threshold for irradiation at shorter wavelengths is expected due to the more efficient photoionization \cite{olivie2008wavelength}. However, in our study, the gradual linear damage at shorter wavelengths, owing to linear absorption in tissue, prevents the irradiated samples from reaching the cavitation regime. In contrast, for irradiation at 1030\,nm in both the absence and presence of a two-photon absorption path of fluorescence proteins, the damage appears abruptly. The cavitation was observed at an average fluence of \SI{1.2}{J/cm^2} at 2\,MHz operation and an average fluence of \SI{2}{J/cm^2} for 127\,kHz operation, which is consistent with previous observations on ablation threshold in porcine corneal stroma \cite{olivie2008wavelength}. The severity of cavitation escalated when the larvae were irradiated at 2\,MHz compared to 127\,kHz. No luminous plasma was observed at these regimes, resembling the optical breakdown in aqueous media when irradiated with femtosecond pulses \cite{venugopalan2002role,hammer1996experimental,noack1999laser, docchio1986experimental}.

Water can be treated as an amorphous semiconductor with a bandgap energy of 6.5\,eV. For a transition from the molecular 1b$^1$ orbital into an excitation band, the energy of six photons at 1030\,nm is required \cite{sacchi1991laser, williams1976liquid}. In our study, we reach the critical peak power of 94\,GW at 127\,kHz and 57\,GW at 2\,MHz for irradiation at 1030\,nm, which are clearly beyond the 6.2\,MW of water critical power, considering the nonlinear index of $1.9\times10^{-20}m^2/W$ \cite{wilkes2009direct}. The energy thresholds for optical breakdown at 50\% breakdown probability for the targeted CNS cells at z $\approx$75\,$\mu$m are 7.8\,nJ at 127\,kHz and 5\,nJ at 2\,MHz extracted from Figure \ref{fig2} a. The sharpness of the nonlinearity of the damage, which is defined by the ratio of the threshold energy to the energy difference between a 10\% and a 90\% breakdown probability, is calculated to be 2.7 at 127\,kHz and 1.8 at 2\,MHz \cite{577307}, emphasizing on the steeper break down at lower repetition rates.

Assuming pure water and according to simulated plasma dynamics and electron energy spectrum using the multi-rate-equation model \cite{liang2019multi}, a 1030\,nm, 127\,kHz, 250\,fs laser pulse at 8.1\,$TW/cm^{2}$ peak intensity produces an electron density of the order of 10$^{21}\,cm^{-3}$, which is similar to the free electron density at the optical breakdown threshold of water \cite{vogel2005mechanisms, liang2022probing, linz2016wavelength}. This regime of nonlinearity leads to 10$^3$\,K temperature rise by a single pulse at 1030\,nm, where only heating by nonlinear absorption is considered \cite{liang2019multi}. For femtosecond laser pulses, heat accumulates if the heat diffusion time is longer than the inter-pulse interval. For samples irradiated at 4.9\,$TW/cm^{2}$ at 2\,MHz repetition rate, the generated free electron density is reaching the low plasma density. The inter-pulse interval at this repetition rate is 500\,ns, which is considerably longer than the heat dissipation time. For focusing at a numerical aperture lower than 0.9, the steady-state temperature distribution is broader than the generated heat due to the single pulse distribution. However, at repetition rates lower than 2\,MHz, temperature accumulation and broadening of the temperature distribution are negligible \cite{vogel2005mechanisms} and can not cause the observed decrease in peak intensity threshold for optical breakdown and cell death at 2\,MHz repetition rates \cite{simanovskii2006cellular}. The lowered cavitation threshold could be due to the low-density plasma generation, which could trigger plasma-mediated chemical effects and nonlinear photochemistry \cite{mao2004dynamics}. At 1030\,nm, four photons are required for DNA excitation with a low cross-section. However, in the presence of free electrons, many different DNA damage mechanisms can be initiated by capturing the electron via DNA base \cite{schmalz2023dissection, boudaiffa2000resonant}. Moreover, free-electron-mediated effects become possible once seed electrons in the conduction band of water are created, scaling with the irradiation wavelength \cite{linz2016wavelength}. Therefore, it can be concluded that the generated low-density plasma in samples irradiated with 1030\,nm pulses at 2\,MHz leads to the creation of ROS by the dissociation of water molecules \cite{nikogosyan1983two, tirlapur2001femtosecond,boudaiffa2000resonant} and DNA damage, lowering the ablation threshold at higher repetition rates \cite{vogel2005mechanisms, huels2003single, boudaiffa2000resonant}.

The radius of the damage size for the CNS of the samples irradiated at 1030\,nm is \SI{40}{\micro\metre} for irradiation at 2\,MHz and \SI{13}{\micro\metre} for irradiation at 127\,kHz at z$\approx$ \SI{75}{\micro\metre}. The irradiation peak intensity at 127\,kHz is above the required threshold for bubble formation by a single laser pulse \cite{vogel2005mechanisms}. Therefore, no collateral damage was observed, and the damage was confined to the region around the laser-induced plasma due to mechanical forces arising from bubble formation and shockwave emission \cite{vogel2008femtosecond,vogel1996shock,vogel1999energy}. For samples irradiated at 2\,MHz repetition rates, collateral damage is observed due to bubble formation based on plasma-mediated accumulative chemical disintegration of biomolecules. Free electrons have picoseconds lifetime and nanometers traveling range and cannot have a seminal contribution to the size of the observed damage site \cite{meesungnoen2002low, kai2016dynamic}. However, the generated damaging "agents" at the focal spot, such as ROS, have a high diffusivity and can travel over \SI{25}{\micro\metre} within less than a second \cite{nikogosyan1983two, van1993diffusion}. Moreover, due to our relatively low numerical aperture, the size of the damage could be affected by the self-focusing and aberration of the irradiation beam \cite{vogel1999influence}, while its expansion is restricted by the surrounding matrix \cite{hutson2007plasma}. Therefore, at the peak intensities reported in this work, the generated free electron also contributes to the damage size before reaching the beam waist \cite{hammer1997shielding, arnold2005streak}.

\section*{Conclusion}

For the successful implementation of cutting-edge nonlinear microscopy techniques for deep-tissue, \textit{in vivo} imaging and designing their use for medical applications, a comprehensive understanding of the non-invasive, optimal operational parameters and constraints of the imaging system is essential. This entails fine-tuning variables like the laser pulses' average power, peak intensity, wavelength, pulse duration, and other factors such as imaging dwell time. In addition, a rigorous assessment of thermal or photon-based damage mechanisms across various irradiation wavelengths is required. Equally important is the distinction between the impact of peak intensity, average power, and dwell time in nonlinear microscopy, as the nonlinear signal scales with peak intensity. At the same time, the signal-to-noise ratio is proportional to the average power of the laser pulse and dwell time, while an in-depth assessment of biological tissue requires longer penetration of light and employment of the focusing optics with a long working distance. Though interconnected, these parameters may initiate various photodamage mechanisms, as shown in this study.

Water, a major component of biological systems, exhibits five main resonances in the near-infrared at \SI{0.76}{\micro\metre}, \SI{0.97}{\micro\metre}, \SI{1.19}{\micro\metre}, \SI{1.45}{\micro\metre}, and \SI{1.94}{\micro\metre}, with an absorption cross section increasing by order of magnitude when scaling between the adjunct resonances \cite{Curcio:51}. With recent advancements in ytterbium lasers operating at 1030\,nm \cite{fattahi2014third, Wang:17, Buberl:16}, alongside novel detection schemes \cite{herbst2022recent}, the second near-infrared window of the tissue has become an increasingly fitting candidate for label-free nonlinear imaging \cite{NIRII, NIRII2}.

When conducting \textit{in vivo} deep-tissue nonlinear microscopy, it is crucial to evaluate various photodamage mechanisms collectively since they operate in unison at differing intensities upon the interaction of femtosecond pulses with dense tissue. Our study of the long-range focusing identifies two different cavitation regimes. The damage is confined and well-controlled at low repetition rates due to plasma-based ablation and sudden local temperature rise. At high repetition rates, the damage becomes collateral due to plasma-mediated photochemistry and is not influenced by the thermal accumulation of consecutive laser pulses. Our findings agree well with the results of a theoretical model developed by Liang et al. \cite{liang2019multi}. Fluorescence labels are employed to visualize photodamage occurring in the tissue and study their influence on the photodamage. While the mKate2 has a two-photon absorption pathway when irradiated with femtosecond pulses at 1030\,nm, GFP does not have a strong nonlinear absorption path. The values shown in Figure \ref{fig2} f demonstrate a negligible influence of the ROS activated by two-photon absorption in mKate2 on photodamage at a low repetition rate of 127\ kHz. Furthermore, our study reveals that at peak intensities exceeding the CNS damage threshold, no discernible damage is evident at the epidermal level. This behavior can be associated with the four times lower peak intensity on the skin than the spinal cord. Moreover, we demonstrate that it is possible to extend the dwell time; therefore, photon flux, extensively and noninvasively for irradiation at 1030\,nm, as long as the peak intensity is below the low-plasma density threshold. In this regime, the associated photon flux to the peak intensity damage threshold can be exceeded by increasing the dwell time without observation of the damage in the sample. This behavior is not observed for irradiation at the harmonics at 515\,nm and 343\,nm, emphasizing the lower scaling of the free electron density versus peak intensity at shorter irradiation wavelengths \cite{liang2019multi}.

These findings significantly contribute to advancements in innovative microscopy techniques, like multimodal microscopy \cite{ guesmi2018dual} or femtosecond fieldoscopy, in which high spatial resolution, label-free images with attosecond temporal resolution can be captured \cite{herbst2022recent, srivastava2023near, fattahi2023method, scheffter2024compressed}. We not only lay down guidelines for emerging nonlinear imaging techniques but also suggest a strategy for inducing highly precise, localized injuries, which is of interest to various research fields such as neuroregeneration \cite{tsata2021switch, matrone2013laser}.

\section*{Methods}

\textbf{Irradiation and live imaging}:
An Yb:KGW amplifier (CARBIDE from Light Conversion) was used to irradiate zebrafish. \SI{20}{\micro J} (on the microscope: from 3.75\,pJ up to 14.2\,nJ) of the output of the laser was frequency doubled in a 1.5\,mm thick BBO crystal with 33.7° type I phase matching angle. A second BBO crystal was used for frequency tripling of the laser via cascaded second-order effects by sum frequency generation between 1030\,nm and 515\,nm pulses \cite{homann2008octave}. The crystal was 1.5\,mm thick with a phase-matching angle of 62.8°, type II. A high reflective mirror (Eksma, 042-515) was used as a beam splitter to separate the fundamental from the second-harmonic signal. Afterward, the second harmonic beam was separated from the third harmonic beam by a harmonic separator (Eksma, 0423535pht). The samples were irradiated at 1030\,nm, 515\,nm and 343\,nm. A reflective objective with a numerical aperture 0.5, magnification of 40, and a focal length of 5\,mm (Thorlabs) was used to focus the beam on the zebrafish spinal cord. The beam size of the three beams was scaled to 5.1\,mm to fill the 5.1\,mm aperture of a reflective objective. The size of the beams at the focus was calculated using Abbe’s resolution limit, resulting in beam sizes of \SI{0.2}{\micro\metre}, \SI{0.3}{\micro\metre}, and \SI{0.6}{\micro\metre} for respective irradiation wavelengths. The irradiation peak intensity was calculated based on these values. As the Rayleigh length of the focused beam is shorter than the transversal length of the samples, a CCD camera is used to ensure the light is focused on the spinal cord. To calculate the peak intensity of the laser pulses on the skin of the zebrafish, it was assumed that the distance from the spinal cord to the skin of 3\,dpf zebrafish is \SI{100}{\micro\metre}. This information was used to calculate the beam size and the corresponding peak intensity of the irradiation beams on the skin. Live examination and imaging before and after irradiation were done using the following commercial microscope setups: 1) Plan-Apochromat 10x/0.45 M27 objective or Plan-Apochromat 20x/0.8 objective on a Zeiss LSM 980 confocal microscope. The detector ranges used were 490-588 nm for EGFP, 597-695 nm for mKate2 and mCherry, and 641-695 nm for Alexa fluor 647. We used a main beam splitter (MBS) 488/594 for the detection of EGFP, mKate2, and mCherry, and an MBS 488/561/639 for the detection of Alexa fluor 647. 2) Plan-Apochromat  1.0x  or 2.0x objective on a Leica M205 FCA stereo microscope equipped with emission filters ET525/50 nm (GFP) and ET630/75 nm (mKate2, mCherry), and a Leica DMC6200 C color camera.  For repetitive imaging, larvae were released from agarose after imaging, transferred to embryo medium at 28.5 °C, and re-mounted the following day. The emission peak for mKate2 is at 633 nm, for mCherry at 610 nm, for EGFP at 507 nm, and for Alexa fluor at 671 nm.

\textbf{Zebrafish husbandry and transgenic lines}: All zebrafish lines were maintained and reared in a research fish facility (Tecniplast, Italy) under a 14/10\,h light/dark cycle according to FELASA recommendations \cite{brand2002keeping,alestrom2020zebrafish}. All experimental procedures were done on larval zebrafish aged up to 120 hours post-fertilization (hpf), derived from voluntarily mating adult zebrafish (permit: I/39/FNC of the Amt für Veterinärwesen und gesundheitlichen Verbraucherschutz Stadt Erlangen). Procedures on zebrafish larvae up to 120 hpf is not regulated as animal experiments by the European Commission Directive 2010/63/EU. For all irradiation experiments, we used 72 hpf zebrafish of either sex. For experimental analyses, we used the following available zebrafish lines: wild-type AB, BAC(\textit{pdgfrb}:Gal4ff)\textsuperscript{ncv24} \cite{ando2016clarification}, 5x\textit{UAS}:EGFP\textsuperscript{zf82} \cite{asakawa2008genetic}, \textit{mpeg1}:mCherry\textsuperscript{gl23} \cite{ellett2011mpeg1}, \textit{elavl3}:rasmKate2\textsuperscript{mps1}\cite{tsata2021switch}, \\\textit{her4.3}:GFP-F\textsuperscript{mps9}\cite{kolb2023small}, \textit{krtt1c19e}:EGFP\textsuperscript{sq1744}\cite{lee2014basal}, and
\textit{elavl3}:GFP-F\textsuperscript{mps10}\cite{kolb2023small}. The combination of \\BAC(\textit{pdgfrb}:Gal4ff) and 5x\textit{UAS}:EGFP transgenic lines were abbreviated as \textit{pdgfrb}:GFP. Embryos were treated with 0.00375\,\% 1-phenyl-2-thiourea (Sigma-Aldrich \#P7629) beginning at 24 hpf to prevent pigmentation.

\textbf{Zebrafish mounting}: Zebrafish larvae were anesthetized in E3 medium containing 0.02\,\% Ethyl 3-aminobenzoate methanesulfonate (MS-222; PharmaQ \#Tricaine PharmaQ) and mounted in a lateral position in 1\,\% low melting point agarose (Ultra-PureTM Low Melting Point, Invitrogen Cat\#16520) between two microscope cover glasses. Larvae were covered with 0.01\,\% MS-222-containing E3 medium to keep preparations from drying out. 

\textbf{Whole-mount TUNEL labelling}:
Terminally anesthetized larvae were fixed in 4\% PFA (Thermo Fisher Scientific Cat\#28908) in PBS overnight at 4°C. After removing the head and tail using micro scissors, larvae were permeabilized by subsequent incubation in acetone and Proteinase K (Invitrogen Cat\#25530-049) as described elsewhere \cite{john2022mechanical}. Samples were re-fixed in 4\% PFA in PBS and Click-iT TUNEL Alexa Fluor 647 Imaging Assay (Thermo Fisher Cat\#C10247) was performed according to the manufacturer’s protocol to label apoptotic cells. Briefly, samples were equilibrated in TdT reaction buffer for 30 min at room temperature, followed by incubation in TdT reaction cocktail overnight at room temperature. Samples were washed in PBTx and mounted in 75\% glycerol in PBS. Imaging was done using a Plan-Apochromat 20x/0.8 objective on a Zeiss LSM 980 confocal microscope.

\section*{Data availability}
The data that support the findings of this study are available from the corresponding authors upon reasonable request. 

\section*{Acknowledgment}
We thank Casandra Cecilia Carrillo Mendez and Olga Stelmakh for excellent fish care. H.F. acknowledges financial support from the Max Planck Society. D.W. acknowledges financial support from the Deutsche Forschungsgemeinschaft (project number 460333672 – CRC1540 Exploring Brain Mechanics (subproject B05)). S.J. acknowledges the Erlangen Graduate School scholarship in Advanced Optical Technologies by the Bavarian State Ministry for Science and Art.

\section*{Author Contributions}
H.F. and D.W. designed the study. S.J. and A.H. performed the measurements. N.J., J.K., and D.W. prepared the samples and performed imaging. S.J., A.H., K.S., D.W., and H.F. performed the data analysis. H.F., D.W., and S.J. wrote the manuscript. All authors reviewed and contributed to the manuscript text.

\section*{Competing interest}
The authors declare no competing interest. 

\printbibliography
\newpage

\end{document}


\begin{center}
{\Large\bfseries Nonlinear dynamics of femtosecond laser interaction with the central nervous system in zebrafish \par}
\vspace{3ex}
{\bfseries
Soyeon Jun$^{1,2,4}$, Andreas Herbst$^{1,2}$, Kilian Scheffter$^{1,2}$, Nora John$^{1,2,3}$, Julia Kolb$^{1,2,3}$, Daniel Wehner$^{1,3}$, Hanieh Fattahi$^{1,2}$\par}
{\footnotesize\itshape
    1. Max Planck Institute for the Science of Light, Staudtstr. 2, 91058 Erlangen, Germany\\
    2. Friedrich-Alexander University Erlangen-Nürnberg, Staudtrstr. 7, 91085 Erlangen, Germany\\
    3. Max-Planck-Zentrum für Physik und Medizin, Staudtstraße 2, 91058 Erlangen, Germany \\
    4. Friedrich-Alexander-Universität Erlangen-Nürnberg (FAU), Erlangen Graduate School in Advanced Optical Technologies (SAOT), Paul-Gordan-Str. 6, 91052 Erlangen, Germany\\
    \par}
\vspace{3ex}
\end{center}

\FloatBarrier
\begin{figure}[h!]
    \centering
    \includegraphics[width=1\textwidth]{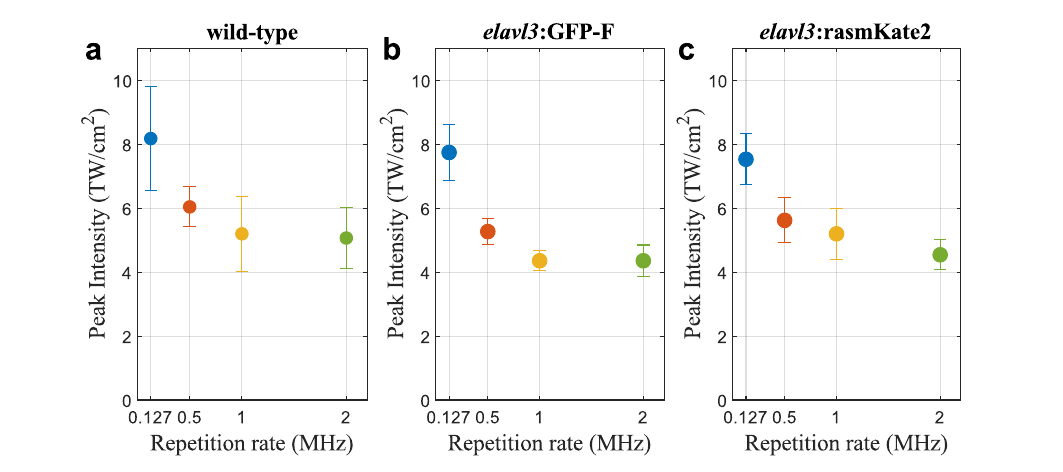}
    \caption{Three types of zebrafish larvae were prepared: two transgenic lines in which neurons were labeled either by mKate2 (assessed in \textit{elavl3}:rasmKate2 transgenic animals) or GFP (assessed in \textit{elavl3}:GFP-F) transgenic animals, and non-labeled wild-type zebrafish for comparison. The measured peak intensity damage thresholds for a) wild-type zebrafish, b) GFP labeled zebrafish and c) mKate2 labeled zebrafish at four different pulse repetition rates of 127\,kHz, 500\,kHz, 1\,MHz, and 2\,MHz.}
    \label{SI:fig1}
\end{figure}
\begin{figure}[h!]
    \centering
    \includegraphics[width=0.25\textwidth]{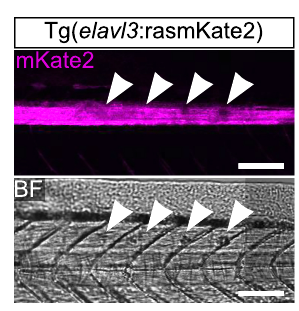}
    \caption{Shown is an \textit{elavl3}:rasmKate2 transgenic zebrafish larva, which was irradiated by 343\,nm laser pulses at an average power below the damage threshold. Arrowheads indicate the focus of irradiation. Increasing the irradiation time from 30 s to 120 s expands the lesion area. Images show fluorescently labeled spinal cord (neurons; mKate2) and brightfield recordings. Scale bars: 50\,$\mu$m.}
    \label{SI:fig2}
\end{figure}
\begin{figure}[h!]
    \centering
    \includegraphics[width=1\textwidth]{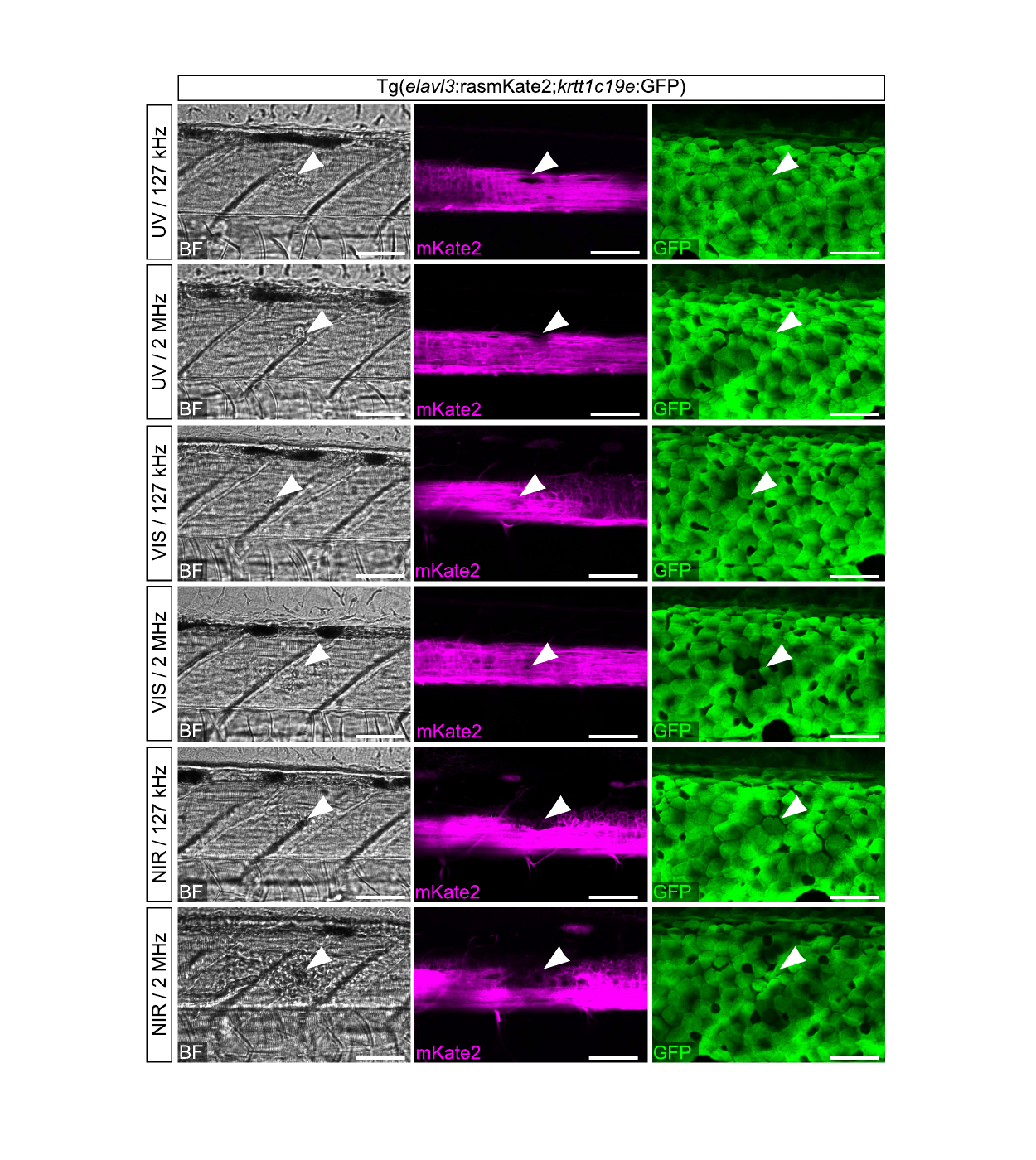}
    \caption{Shown are trunks of a transgenic zebrafish larva with fluorescently labeled spinal cord (neurons; mKate2) and epidermis (basal keratinocytes; GFP) after 30\,s irradiation with 1030\,nm, 515\,nm, and 343\,nm pulses above the damage threshold. Arrowheads indicate the focus of irradiation. While damage is visible at the level of the spinal cord, the superficially located epidermis stays intact. Images shown are bright field (BF) recordings, single optical sections (mKate2), or orthogonal projections (GFP) of the confocal image. UV: Ultra-violet, VIS: visible, and NIR: Near-infrared Scale bars: 50\,$\mu$m.}
    \label{SI:fig3}
\end{figure}
  \printbibliography